\documentclass{ws-procs9x6}
\usepackage{amssymb}
\begin{document}

\title{Thermal features far from equilibrium: Prethermalization}
\author{Szabolcs Bors\'anyi}
\address{Institute of Theoretical Physics, University of Heidelberg,\\ 
Philosophenweg 16, Heidelberg 69120, Germany\\
}
 
\maketitle

\abstracts{
The phenomenon of prethermalization and the subsequent steps
of thermalization are analyzed in the framework of the chiral
quark model. We solve the quantum equations of motion of the
field theory derived from the 2PI effective action and study
the time scales of equilibration. We find that already
after a 0.6 fm/c long period of time some equilibrium features
appear, even though the system is still far from equilibrium.
This might be an ingredient for
understanding the success of ideal hydrodynamic description.}

 

\section{Introduction}

One of the crucial issues in heavy ion physics is the early
thermalization of the excited quark and gluon fields:
The extreme success of hydrodynamic calculations started earlier than 1 fm/c
require the theoretical establishment
of the early existence of an equation of state.\cite{kolb_sewm}

The proper description of equilibration is a longstanding challenge in
particle physics. Kinetic theories are popular tools for studying
the series of incoherent collisions. By neglecting the off-shell processes
the easiest way is to follow the elastic collisions only, but these
deliver thermalization times by orders of magnitudes larger than desired.
Equilibration may be enhanced by inelastic scatterings so that it
completes before chemical freeze-out\cite{bottomup}.
This way, thermalization times reduce to 2-3 fm/c, which is still longer
than the early use of flow equations would require\cite{julien}.

Recent field theoretical developments enable the nonequilibrium treatment
of quantum fields by performing a systematic ladder resummation\cite{bergesandco}. The available numerical techniques and resources
enable us to solve the time evolution of 3+1 dimensional quantum fields
in a self-consistent approximation scheme. These methods are not limited
to scalar fields any more, nowadays the 3+1 dimensional chiral quark model
(with two quark flavors and 4 scalars) may also be routinely solved\cite{fermions}. 

Thermalization time is not a single number, of course. Different
degrees of thermalization (LTE, kinetic and chemical equilibration)
can have very different time scales ranging from the rapid convergence
of bulk quantities to the slow equilibration of energy spectra\cite{preth}.

\begin{figure}[t]
\begin{center}
\hbox{
\parbox[b]{5.6cm}{\epsfxsize=5.6cm \epsfbox{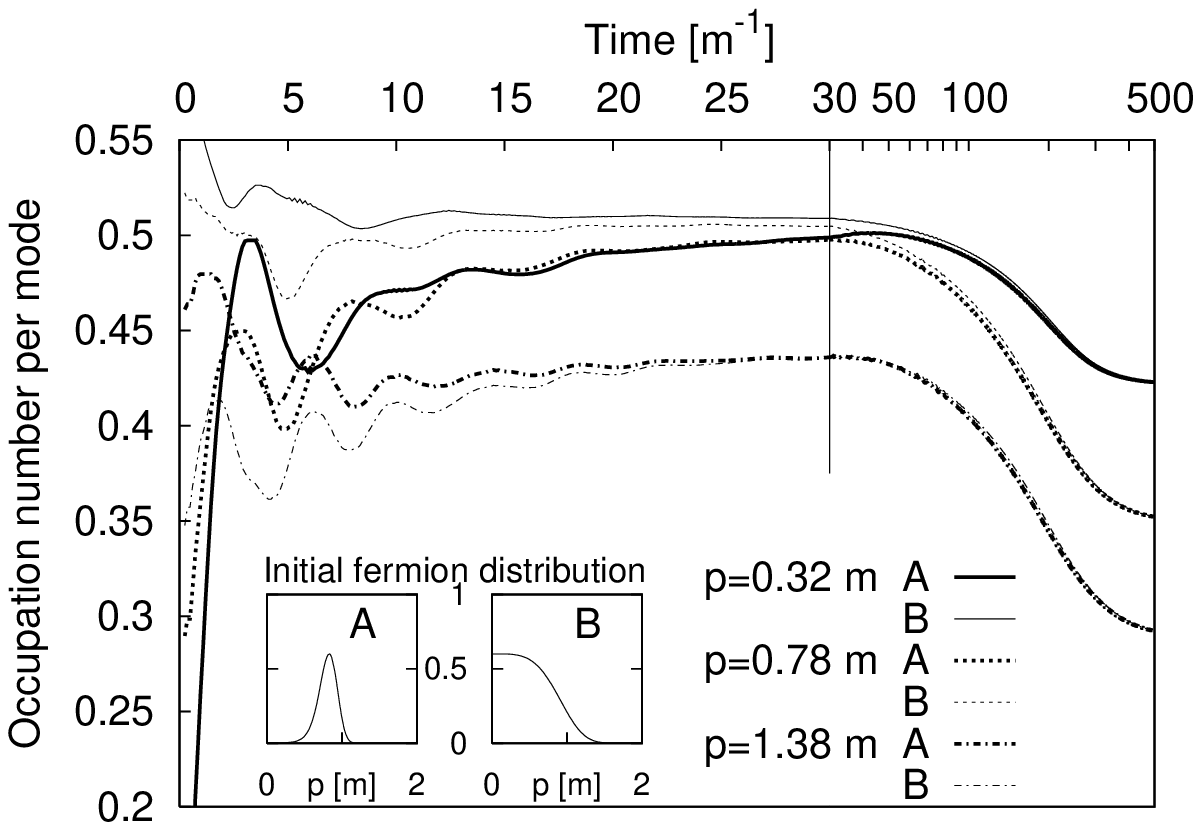}\vspace*{3.6mm}}
\parbox[b]{5.6cm}{\epsfxsize=5.6cm \epsfbox{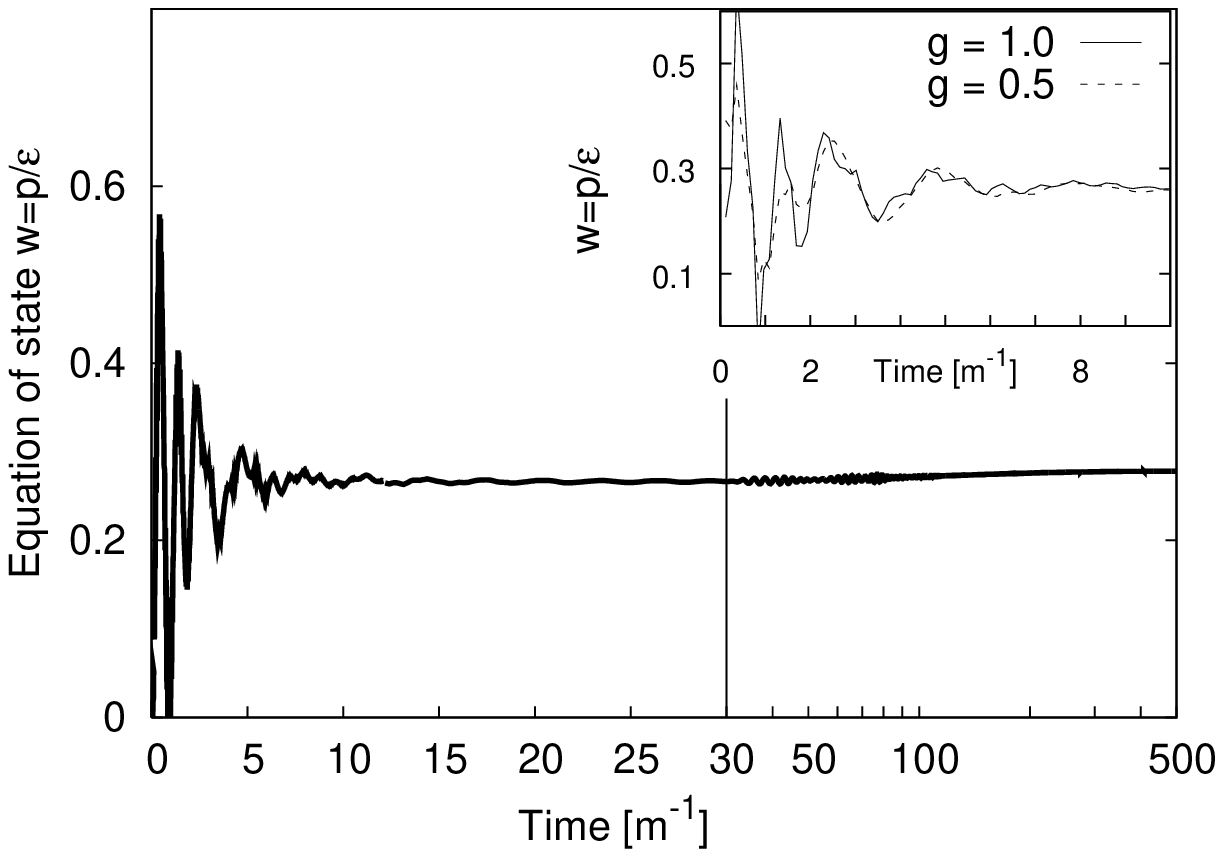}}
}
\end{center}
\vspace*{-8mm}
\caption{{\em Left:} The fermionic occupation number $n^{(f)}(t;p)$
for three different momentum modes as a function of time. 
The evolution is shown for two different initial
conditions with {\sl same} energy density. The dynamics 
becomes rather quickly insensitive to the initial 
distributions (A) and (B) displayed in the insets -- much before the modes 
settle to their final values.
{\em Right: }
The ratio of pressure and energy density as a function of
time. One observes that an approximately time-independent equilibrium
form of the equation of state builds up very early. The inset
demonstrates for two different couplings that this prethermalization
time is independent of the interaction details. The insensitivity
of bulk quantities to the late-time thermalization scale is in
sharp contrast to mode-by-mode quantities.
The long-time behavior is shown on a 
logarithmic scale for $t \ge 30\, m^{-1}$.}
\label{fig1}
\vspace*{-2mm}
\end{figure}

The model we consider is built up from two massless fermionic degrees of
freedom and four scalars interacting according to the following Lagrangian:
\centerline{$
{ L} =
\Big\{\bar{\psi} i \partial\!\!\!\slash \psi 
+\frac{1}{2}\left[(\partial\sigma)^2+(\partial\pi)^2\right]
+\, \frac{g}{N_f} \bar{\psi} \left[\sigma + i\gamma_5 \tau^a \pi^a \right] \psi
- V(\sigma^2 + \pi^2) \Big\},
$}
where
$\tau^a$ denote the standard Pauli
matrices. We consider a quartic scalar self-interaction $V(\sigma^2 + \pi^2) =
m^2 \left(\sigma^2 + \pi^2\right)/2 + \lambda \left(\sigma^2+
\pi^2\right)^2/(4! N_f^2)$.  The employed couplings are taken to be 
$g=1$ and $\lambda=1$.  Though these couplings
are smaller than heavy ion applications would require, we will be able to draw
conclusions for the physical values since our main result will turn out to
be coupling independent. The smaller-than-natural coupling leads to a
spectacular separation of the various time scales.

We use the two-particle irreducible (2PI) effective action to two-loop order,
which includes direct scattering as well as off-shell and memory
effects\cite{fermions}. We numerically solve the nonequilibrium gap equation
and obtain the evolution of the propagator $G(t_1,t_2,|\vec x_1 -\vec x_2 |)$
for each degree of freedom. To simplify our equations we assume 
homogeneity and isotropy in space.

\section{Time scales of thermalization}

Thermalization is loss of initial information. Our dynamics is conservative,
still, virtually, we loose information if we do not read out all the internal
variables but only the time-local propagator
$G(t_1\approx t_2,|\vec x_1-\vec x_2|)$. This time-local propagator stores
the information on the particle number distribution\cite{fermions}.
The evolution of these occupation numbers for three different momenta
is shown in Fig.~\ref{fig1}left. We also show the convergence for two different
initial spectra with equal energy density.
Although the details of the initial distribution are mainly washed out
in the rather early evolution ($15|m|^{-1}$), the spectrum is still far
from equilibrium. It takes by an order of magnitude longer to approach
the thermal spectrum (for small coupling).

\begin{figure}[t]
\begin{center}
\hbox{
\epsfxsize=5.6cm\epsfbox{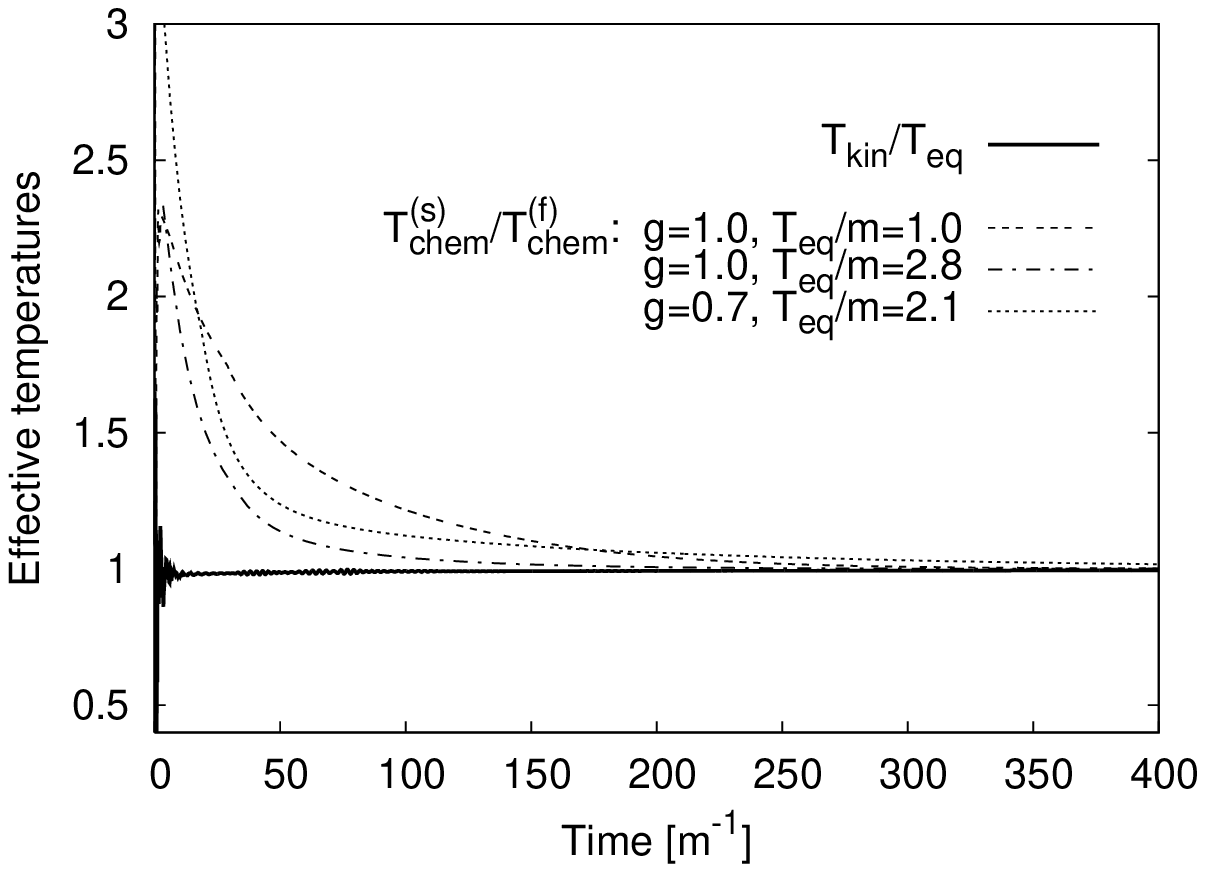}
\epsfxsize=5.6cm\epsfbox{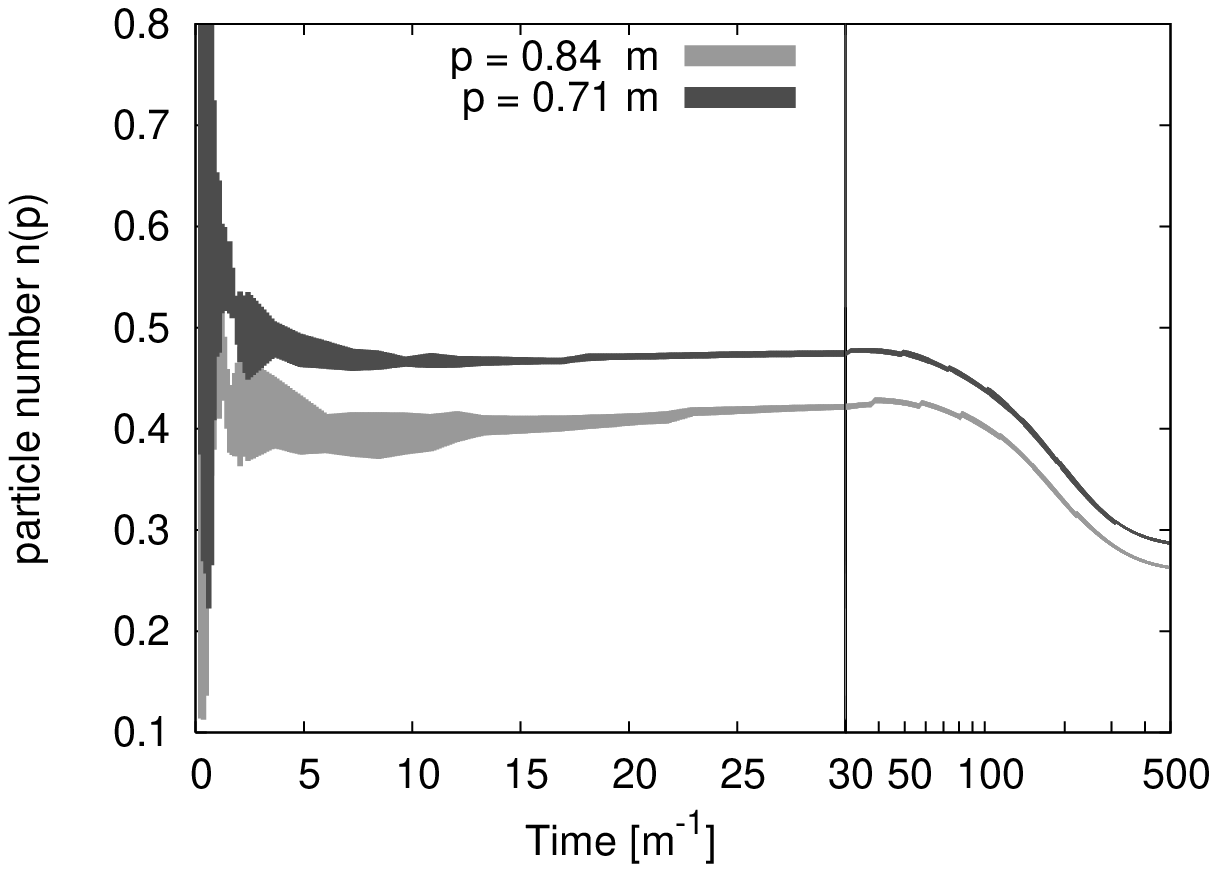}
}
\end{center}
\vspace*{-6mm}
\caption{{\it Left:}
The solid line shows the kinetic temperature $T_{\rm kin}(t)$
as a function of time. In contrast to the mode-by-mode quantities,
the kinetic temperature, which includes all momentum 
modes, exhibits the phenomenon of prethermalization. 
Also shown is the ratio of the scalar and
the fermion chemical temperatures,
$T_{\rm chem}^{(s)}(t)/T_{\rm chem}^{(f)}(t)$, for different 
values of the coupling $g$ and $T_{\rm eq}$. One observes 
hat chemical equilibration does not happen on the prethermalization 
time scale.
{\it Right:}
On the damping time scale a non-equilibrium variant of the
fluctuation--dissipation relation is dynamically realized. This
gives a particle number definition even for strongly coupled systems.
The grey bands demonstrate the particle numbers and the consistency of
their determination.
}
\label{fig2}
\end{figure}

We may split the observed time scales to three epoches. The separation of
these scales we find for perturbative couplings only, in a strong coupling
application the first two might coincide. (The time values correspond to
the parameters used in the presented plots.)

\begin{enumerate}
\item \textit{Prethermalization} --- loss of coherence ($t<5|m|^{-1}$)\par
Prethermalization is a universal far-from-equilibrium phenomenon, its
time scale is given by the relevant scales of the dynamics (temperature, mass)
but it is independent of the coupling, as shown in Fig.~\ref{fig1}right.\par
Contrary to the mode-by-mode occupation numbers (Fig.~\ref{fig1}left) the bulk,
averaged quantities are often insensitive to the details of the spectrum
as it is also observed in classical field theory~\cite{denescikk}.\par
We find that this short time is enough for very rapid establishment of an
almost constant ratio of pressure over energy density (Fig.~\ref{fig1}right),
as well as a kinetic temperature based on average kinetic energy
(Fig.~\ref{fig2}left, for definitions see Ref.\cite{preth}).
These are important ingredients
of the early equation of state. This level of equilibration is one of the
important preconditions of the use of hydrodynamic equations.\par

\item \textit{Damping} --- decay of individual modes ($t<15 |m|^{-1}$)\par
The individual excitations are washed out so that a smooth but still
far-from-continuum spectrum is formed.  We find that this relaxation time
agrees with the inverse width ($\gamma(\vec p;t)$)
of the nonequilibrium spectral function\cite{fermions}.\par
We know that the damping
time (obtained from the imaginary part of the self energy or from the width of
the spectral function) is the decay rate of an individual signal excited over
the ensemble\cite{szepzsolt}. This relaxation time may be much shorter than
thermalization time\cite{aartswetterich}. In the leading order perturbation
theory relaxation time is proportional to the coupling-squared.  We also see
that this relaxation time scale is not a subject of spectacular evolution.
\par
Within this time scale the generically independent symmetric propagator ($F$)
and spectral function ($\rho$) evolve to fulfill a relation (in Wigner
coordinates): $F(\omega,\vec p;t)=\pm(n(\omega;t)+\frac12)\rho(\omega,\vec
p;t)$, which is known as fluctuation--dissipation relation in equilibrium. Here
the nontrivial issue is the momentum ($\vec p$) independence of $n(\omega;t)$.
This enables us to translate it to $n(\vec p;t)$ using the dispersion relation
defined by the peak of the spectral function. The existence of the relation
above provides a particle number definition $n(\vec p;t)$ not tainted by 
quasi-particle picture.\cite{fludi}
Fig~\ref{fig2}right shows by shaded bands how
consistently the particle number $n(\vec p;t)$ is determined from
the evolution of different $\vec k$ modes with
$\omega_k\approx \omega_p\pm\gamma_p$.

\item \textit{Thermalization} --- loss of spectral information
($t\lesssim 500 |m|^{-1}$)\par
Kinetic equilibration is a result of the collective evolution of the particle
spectra.  At the same time the particle abundances in the various species are
equilibrated as well. Whether this longer epoch fits into the short
time range before chemical freeze-out is still a matter of discussion.
\end{enumerate}

\section{Discussion}

Direct heavy ion applications require higher couplings than in the example
above. Tuning the coupling to $g=2.5-3$ one observes that both
relaxation and thermalization time shrinks considerably, approaching the
prethermalization time. The elementary oscillators are then over-damped,
and the estimated prethermalization time will also give a hint for the
damping time scale.  With these couplings one can easily obtain thermal
spectra in the fermionic sector within $1-3$~fm/$c$. 

We find that prethermalization time can be estimated
by the inverse average kinetic temperature\cite{preth}. In particular, we
obtained the following quantitative relation: $T\,t_{pt}\approx 2-2.5$. If one
attempts to use the concept of prethermalization for the case
of heavy ion collisions one has to replace this temperature by the relevant
scale of the dynamics. This we assume to be the saturation scale:
$Q_s\approx1~\textrm{GeV}$ at RHIC energies. Estimating the
prethermalization time by 3 times $1/Q_s$ one has:
$t_{\rm prethermalization}\approx 0.6 \textrm{fm}/c$
This is the very close to the instant that is usually taken as initial
time for the hydrodynamic evolution\cite{kolb,hirano}.

If the complete local equilibration cannot be established, it is worth thinking
if this partial equilibration can already establish the use of flow equations.
We do believe that the study of partial equilibration can help us overcome the
discrepancy between the long thermalization time predicted by kinetic theories
and the short times used by hydrodynamic models. 

The author acknowledges the fruitful discussions and collaborations on this work
with J.~Berges, J.~Serreau and C.~Wetterich.

\vfill\eject

\begin{thebibliography}{9}
%
%

\bibitem{kolb_sewm}
P.~F.~Kolb,
arXiv:nucl-th/0407066,
contribution to this meeting.

\bibitem{bottomup}
R.~Baier, A.~H.~Mueller, D.~Schiff and D.~T.~Son, Phys. Lett. {\bf B502} (2001) 51; Phys. Lett. {\bf 539} (2002) 46.

\bibitem{julien}
J.~Serreau, Nucl. Phys. {\bf A715}, 805c (2003)

\bibitem{bergesandco}
For a recent review see J.~Berges and J.~Serreau,
arXiv:hep-ph/0302210.

\bibitem{fermions}
J.~Berges, S.~Borsanyi and J.~Serreau,
Nucl.\ Phys.\ B {\bf 660} (2003) 51


\bibitem{preth}
J.~Berges, S.~Borsanyi and C.~Wetterich,
``Prethermalization,''
arXiv:hep-ph/0403234, to appear in Phys.~Rev.~Lett.

\bibitem{denescikk}
S.~Borsanyi, A.~Patkos and D.~Sexty,
Phys.\ Rev.\ D {\bf 68} (2003) 063512


\bibitem{szepzsolt}
S.~Borsanyi and Z.~Szep,
Phys.\ Lett.\ B {\bf 508} (2001) 109

\bibitem{aartswetterich}
G.~Aarts, G.~F.~Bonini and C.~Wetterich,
Nucl.\ Phys.\ B {\bf 587} (2000) 403

\bibitem{fludi}
J.~Berges, Sz.~Bors\'anyi C.~Wetterich, in preparation

\bibitem{kolb}
P.~F.~Kolb, J.~Sollfrank and U.~W.~Heinz,
Phys.\ Rev.\ C {\bf 62} (2000) 054909

\bibitem{hirano}
T.~Hirano and Y.~Nara,
arXiv:nucl-th/0404039; 

\end{thebibliography}
\end{document}